\documentclass[11pt]{article}
\usepackage{graphicx}
\usepackage{cs12,html,hyperref}

\begin{document}
 
\title{Infrared Colors of L and T Dwarfs}
 
\author{S.~K.\ Leggett\altaffilmark{1}, D.~A.\ Golimowski\altaffilmark{2}, 
X.\ Fan\altaffilmark{3}, T.~R.\ Geballe\altaffilmark{4} and \\
G.~R.\ Knapp
\altaffilmark{5}}
\altaffiltext{1}{Joint Astronomy Centre Hawaii}
\altaffiltext{2}{The Johns Hopkins University}
\altaffiltext{3}{Institute for Advanced Study Princeton}
\altaffiltext{4}{Gemini Observatory Hawaii}
\altaffiltext{5}{Princeton University Observatory}

\index{Late type stars}
\index{Low-mass stars}
\index{Brown dwarfs}
\index{Infrared  observations}

\begin{abstract}
We discuss the behaviour of the $JHKL^{\prime}M^{\prime}$ colors of  L and T~dwarfs,
based on new photometric and spectroscopic data obtained at the United Kingdom 
Infrared Telescope in Hawaii.   We have measured the first accurate $M^{\prime}$  
photometry for L and T dwarfs.  The $K$--$M^{\prime}$ colors of T dwarfs are much 
bluer than predicted by published models, suggesting that CO may be more abundant 
than expected, as has been found spectroscopically for the T6 dwarf Gl~229B.  We also find that 
$K$--$L^{\prime}$ increases monotonically through most of the M, L, and T subclasses, 
but it is approximately constant between types L6 and T5, due to the onset of CH$_4$ 
absorption at the blue edge of the $L^{\prime}$ bandpass.  The $JHK$ colors of L 
dwarfs show significant scatter, suggesting variations in the amount and properties 
of photospheric dust, and indicating that it may not be possible to associate a 
unique T$_{\rm eff}$ with a given L spectral type.  The $H-K$ colors of
the later T dwarfs also show some scatter which we suggest is due to variations in 
pressure--induced H$_2$ opacity, which is sensitive to gravity (or age for a brown 
dwarf) and metallicity.

\end{abstract}

\section{Introduction and Photometric Systems}

This work is based on  papers by Leggett et al. 
(2001) and  Geballe et al. (2001a).   Geballe et al.
present a  spectral classification scheme for L and T dwarfs 
(summarised elsewhere in these proceedings) and Leggett et al.
present $ZJHKL^{\prime}M^{\prime}$ photometry of a sample of 58 
late--M, L, and T dwarfs.  The L and T dwarfs in these papers are 
primarily taken from recent results of red and near--infrared imaging 
sky surveys --- the Deep Near--Infrared Survey (DENIS,
e.g. Mart\'{\i}n et al. 1999), the 2 Micron All--Sky Survey
(2MASS, e.g. Kirkpatrick et al. 1999) and the Sloan Digital Sky 
Survey (SDSS, e.g. York et al. 2000).
The data presented in this work were obtained at the 
United Kingdom Infrared Telescope (UKIRT) in Hawaii.

Figure 1 shows $J - H$, $J - K$ and $K - L^{\prime}$  colors 
as a function of spectral type, where the type is taken from Geballe 
et al. (2001a).  Section 2 discusses the 1---2.5~$\mu$m colors
of L and T dwarfs ---
$J$ (1.25~$\mu$m), $H$ (1.65~$\mu$m) and  $K$ (2.2~$\mu$m).
Section 3  discusses the longer wavelength 3---5~$\mu$m colors ---
$L^{\prime}$ (3.8~$\mu$m) and $M^{\prime}$ (4.7~$\mu$m). 

\begin{figure}[h!]
\hspace{1cm}
\includegraphics[angle=-90,width=18cm]
{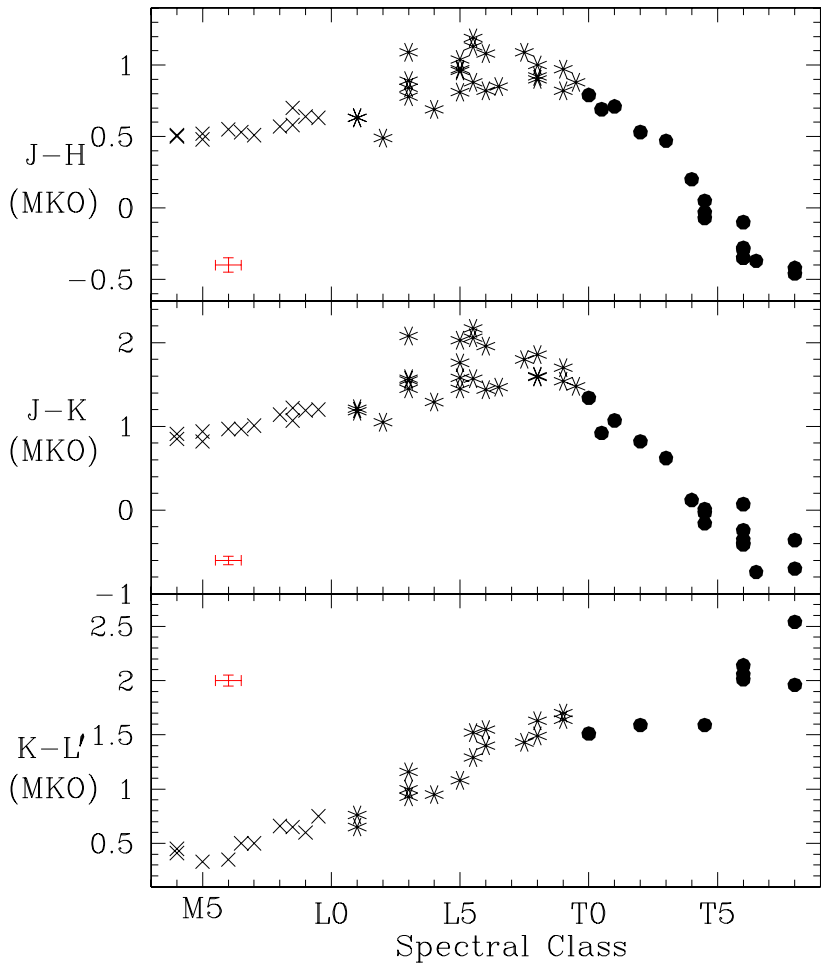}
\caption{$J - H$, $J - K$ and $K - L^{\prime}$ against spectral type
for M dwarfs (crosses), L dwarfs (asterisks) and T dwarfs (filled circles);
photometry from Leggett et al. 2001, spectral classification by 
Geballe et al. 2001a.  Typical error bars are shown in red.}
\end{figure}
 
The infrared passbands are constrained by the transmission of the
terrestrial atmosphere.  Unfortunately, historically,
there has not been a consensus on the exact specifications
of the filters used by different observatories.  Recently a group led by 
\htmladdnormallink{A. Tokunaga}{http://www.ifa.hawaii.edu/~tokunaga/filterSpecs.html}
has defined a set of infrared filters well matched to the atmosphere
(Simons \& Tokunaga 2002, Tokunaga \& Simons 2002);
these filters are being widely adopted and are known as the Mauna Kea 
Observatory (MKO) filter set.  The colors in Figure 1 are on this system.
Note that differences between colors measured on different photometric systems 
can be significant, especially in the case of T dwarfs,
and colors available in the literature should not
be compared without transforming between systems.  This is discussed further in
Sections 2 and 3.

\section{1---2.5~$\mu$m Colors ($JHK$)}

Figure 2 shows the observed 1---2.5~$\mu$m flux distributions of an L and a T dwarf,
with the important absorbing molecular species indicated along the top 
of the figure (see also the paper by Burgasser et al. in these proceedings,
and Burgasser et al. 2001).
We have overlaid $J$, $H$ and $K$ filter profiles from three photometric systems: 
2MASS, UKIRT (prior to adoption of the MKO filters)
and MKO.  Note the significant differences between the filters
and the very structured energy distributions within the bandpasses,
especially for the T dwarfs.  While differences between measured $JHK$
magnitudes on these systems are only around 5\% for L dwarfs, for T dwarfs the 
difference at $K$ is $\sim$10\% and at $J$ it is $\sim$30\%.

\begin{figure}[h!]
\includegraphics[angle=-90,width=12cm]
{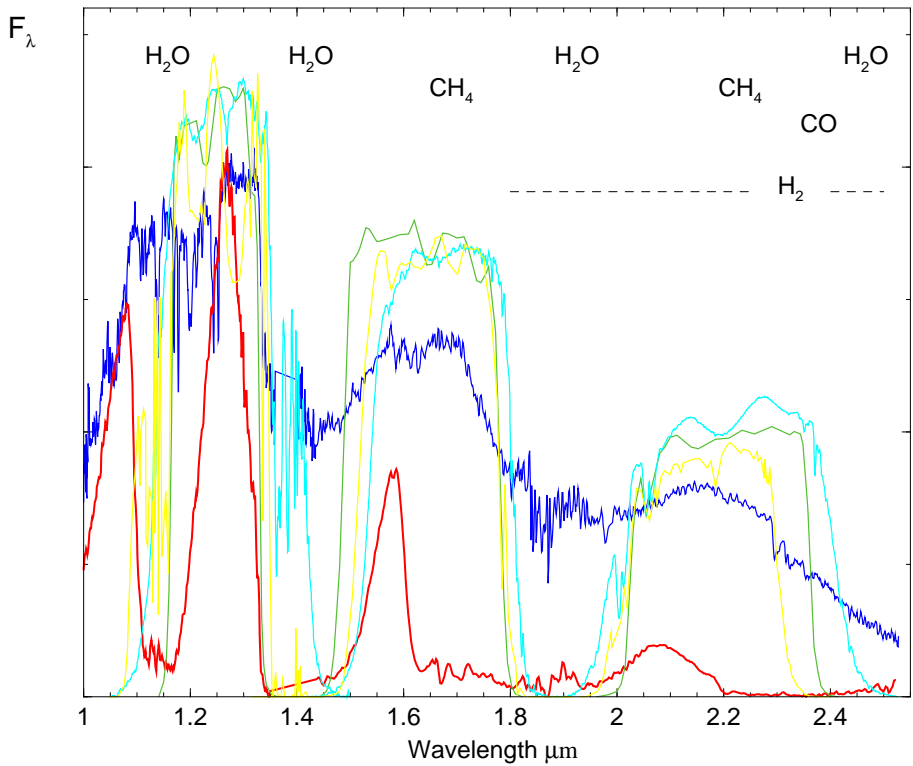}
\caption{Observed spectra of an L1 dwarf (2MASSI J0746+20, Reid et al. 2001, blue line)
and a T8 dwarf (Gl 570D, Geballe et al. 2001b, red line). $J$,$H$,$K$ filter profiles
are overlaid from the MKO (green), UKIRT (cyan) and 2MASS (yellow) photometric
systems.}
\end{figure}

Methane (CH$_4$) absorption in the K band becomes important for spectral
types L8 and later, and in the H band for types T0 and later
(Geballe et al. 2001a).  Hence the $J - H$ and $H - K$ colors become 
increasingly blue for the latest spectral types, as can be seen  in Figure
1.   Figure 1 also shows that while trends are apparent between color 
and type, for the L3---L8 dwarfs there is a large scatter in 
$J - H$ and $J - K$ as a function of type. The scatter is much larger than
the measurement error, indicated by the error bar in the figure.  

\begin{figure}[h!]
\includegraphics[angle=-90,width=12cm]
{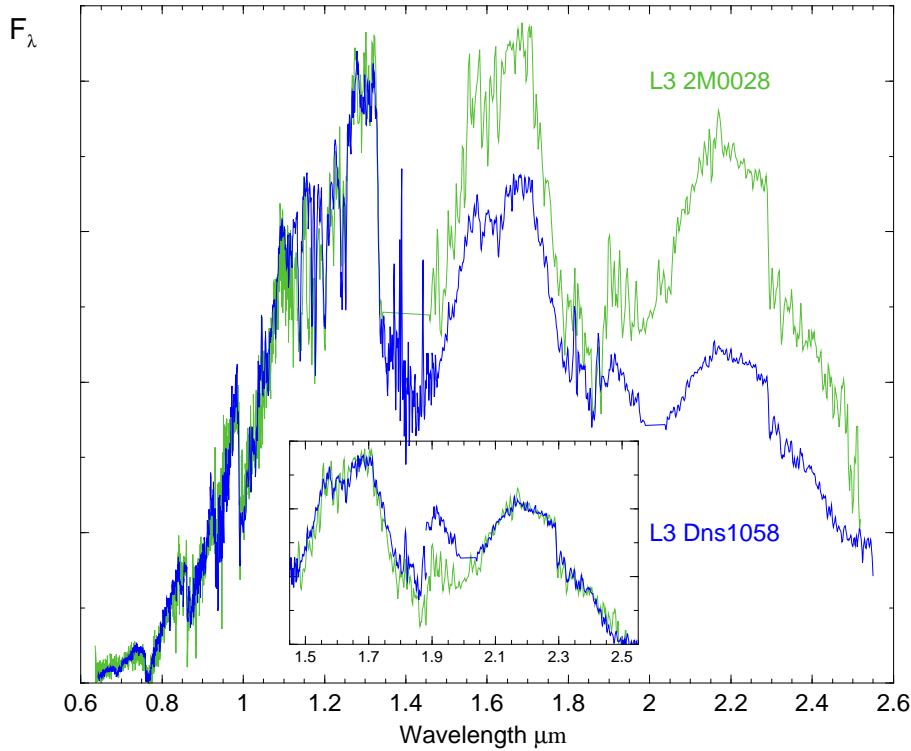}
\caption{Observed relative flux distributions of two L3 dwarfs, from Geballe
et al. 2001a; inset shows different scalings, see text.}
\end{figure}

Figure 3 explores this observed scatter further by superimposing the energy
distributions of two objects both
classified as L3; it can be seen immediately that  they have very different
$J - K$ color.  However the inset shows that if the H and K spectral segments are 
separately  scaled and overlaid, the agreement is excellent.  That is,
the slope of the red continuum, the wings of the water (H$_2$O) bands, and the 
depth of the carbon monoxide (CO) band all match and so indeed the objects are both
L3 types.  The difference in color is probably due to another important
opacity source not yet mentioned: dust.  At the temperatures of these 
photospheres, around 2000~K, grains are expected to form.  This is
discussed further by Marley et al. in these proceedings (see also
Ackerman \& Marley 2001 and Allard et al. 2001). This dust is
expected to cause a warming of the photosphere and a redistribution of
flux to the infrared.  The scatter in color may indicate varying dust
properties caused by differences in metallicity (which affects dust
abundance), age (which limits settling time) and rotational velocity
(which may inhibit dust settling).  Detailed models are required
before we can understand how dust can change the overall color of an L
dwarf without affecting the dominant spectroscopic features.  

Note also how very different the radiated energy is for these two L3 dwarfs
over this wavelength range.  One effect of the dust may be 
that a unique effective temperature cannot be associated with a given L 
spectral type.  Bolometric magnitudes are needed to investigate this further.

Figure 1 shows there is also some scatter in the $J - K$ colors of the 
later T dwarfs.  Figure 4 shows the relative flux distributions of two T8 
dwarfs superimposed; the inset shows the H and K regions scaled separately.
Again it can be seen that although the overall color is different, the
H$_2$O and CH$_4$ absorptions match well, and they are both therefore classified 
as T8.  At the effective temperatures of these atmospheres, around 1000~K, 
grains are expected to lie below the photosphere, and the absorption bands of 
CH$_4$ and H$_2$O are close to saturation.  One opacity source that is
still important is pressure--induced molecular hydrogen (H$_2$) absorption, 
which is an extremely broad feature that spans both the H and K bands but is
strongest at K (see e.g. Borysow et al. 1997).  The strength of this opacity 
is sensitive to gravity (or age, for a brown dwarf) and metallicity, and we 
suggest that the scatter seen in the K band colors of late T dwarfs is due to 
variations in these parameters.

\begin{figure}[h!]
\includegraphics[angle=-90,width=12cm]
{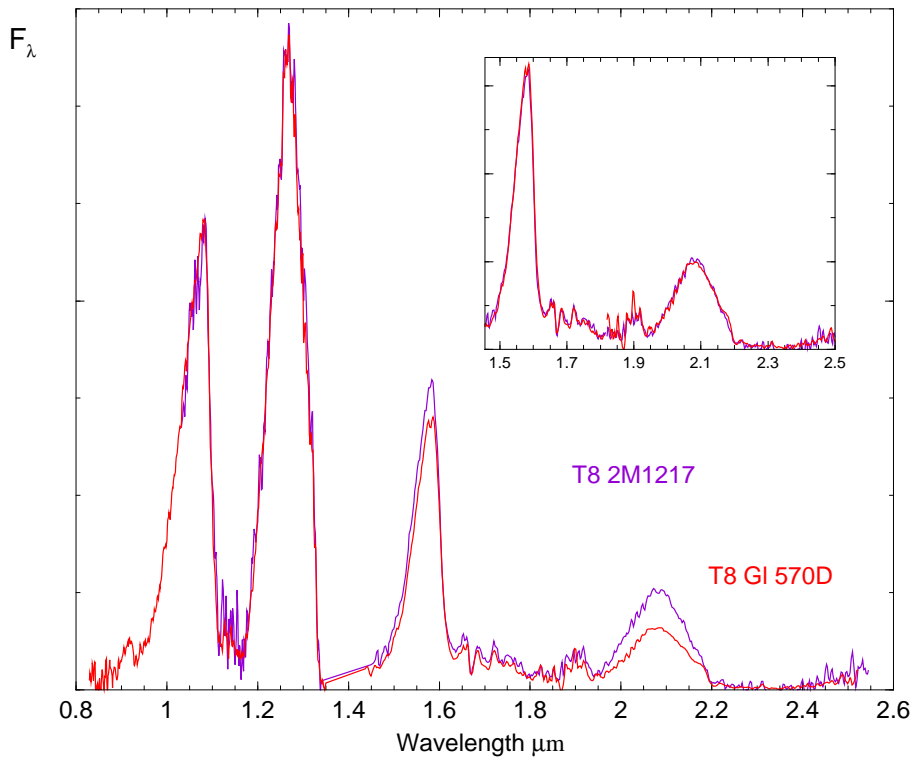}
\caption{Observed relative flux distributions of two T8 dwarfs, from Geballe
et al. 2001a; inset shows different scalings, see text.}
\end{figure}
 
\section{3---5~$\mu$m Colors ($LM$)}

Figure 5 shows the 3---5.5~$\mu$m calculated energy distribution for 
two  model atmospheres with effective temperatures appropriate for an L
dwarf and a T dwarf.  The observed flux distribution for a T dwarf
is also shown.
The principal absorbing species are indicated along the top of the figure and
filter profiles are overlaid.  The MKO $L^{\prime}$ and $M^{\prime}$
filter bandpasses are shown, as well as the earlier UKIRT $L^{\prime}$
profile and a wider $M$ profile.

\begin{figure}[h!]
\includegraphics[angle=-90,width=12cm]
{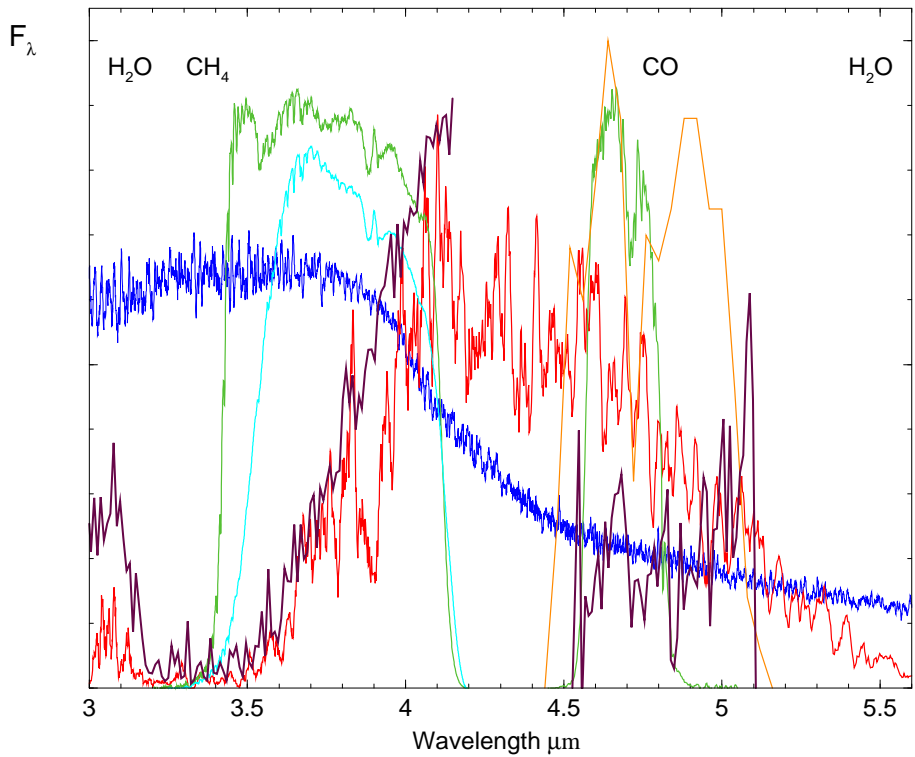}
\caption{Calculated spectra for an L dwarf type atmosphere with T$_{\rm eff} =$ 
2000~K (Allard et al. 2001, blue line), and for a T dwarf
type atmosphere with  T$_{\rm eff} =$  800~K (model by D. Saumon and 
M. Marley, as presented in Geballe et al. 2001b, red line).  Also shown 
is an observed spectrum for the T6 dwarf Gl~229B (Oppenheimer et al. 
1998 and Noll et al. 1997, maroon).  UKIRT (cyan) $L^{\prime}$, 
MKO (green) $L^{\prime}$ and  $M^{\prime}$, and  AAO (orange) $M$
filter profiles are overlaid.
}
\end{figure}
 
In the L band the important opacity sources are H$_2$O and CH$_4$.
Noll et al. (2000) have shown that the 3.3~$\mu$m band of CH$_4$ 
is detected for spectral types around L5 and later.  Figure 1 shows that
our $K - L^{\prime}$  colors are almost constant for types L6 to T5 most 
probably due to the onset of this absorption feature at the blue edge of
our $L^{\prime}$ bandpass.  For later types the band
is saturated and  $K - L^{\prime}$ can increase again.  Differences in the 
L$^{\prime}$  bandpass
can lead to magnitudes that differ by 20\% for T dwarfs (see also
Stephens et al., these proceedings and Stephens et al. 2001).  

In the M band it can be seen that the model spectrum (red line) does not
agree with the observed spectrum (maroon line). Noll et al. (1997)
show that the shape of the observed Gl~229B spectrum indicates the
presence of the fundamental vibration--rotation band of CO 
in this T6 dwarf.  This is unexpected as at these temperatures all the
carbon is expected to be in the form of CH$_4$.  Possibly CO is
being dredged up from hotter, deeper, layers of the atmosphere.

\begin{deluxetable}{lrlrr}
\tablecaption{$K-L^{\prime}$, $K-M^{\prime}$: Model Comparison }
\tablehead{
\colhead{Type} &  \colhead{$\sim T _{\rm eff}$} & \colhead{Observed}
& \multicolumn{2}{c}{Calculated}  \nl
\colhead{} &  \colhead{K}& \colhead{Color} & \colhead{Dusty\tablenotemark{a}} &
\colhead{Settled\tablenotemark{b}}  \nl
}
\tablecolumns{5}
\startdata 
%\nl
 & & $K - L^{\prime}$ & & \nl  
%\nl
L1 & 2100 & 0.7 & 1.0   & \nodata    \nl
L8 & 1400 & 1.6 & 1.8   & \nodata    \nl
T0 & 1300 & 1.5 & 2.1   & \nodata   \nl 
T6 &  950 & 2.0 & 3.3   &     2.3  \nl
\nl
 & & $K - M^{\prime}$ & & \nl   
%\nl
L4   &   1800        & 0.7$\pm$0.1 &  1.0 & \nodata  \nl 
L8   &   1400        & 1.4$\pm$0.1 &  2.0 & \nodata  \nl 
T2   &   1300---1000 & 1.2$\pm$0.2 &  2.1---3.3 & \nodata \nl
T4.5 &   1300---1000 & 1.6$\pm$0.2 &  2.1---3.3 &  $\sim$3.0 \nl
%\nl
\tablenotetext{a}{Chabrier et al. (2000) for ages 
0.1---10~Gyr, corresponding to log~$g\approx$4.2---5.4}
\tablenotetext{b}{Burrows et al. (1997) for  log~$g=$4.5---5.0 
corresponding to ages $\approx$0.3---1~Gyr}
\enddata
\end{deluxetable}

Leggett et al. (2001) give  $K - L^{\prime}$ colors for nineteen L dwarfs and 
eight T dwarfs and $K - M^{\prime}$  colors for two L dwarfs and two T dwarfs.
Table 1 presents these colors as a function of spectral type, where the
$K - L^{\prime}$ color has been averaged over type.  The effective temperature
for each type is shown, estimated using luminosity arguments as described by
Leggett et al.  The colors are compared to those 
calculated by two different models.  The models differ primarily in their
treatment of grain condensation: one has the grains distributed through
the photosphere (the Dusty model, Chabrier et al. 2000), the other
has them below the photosphere so that they do not contribute to the opacity
(the Settled model, Burrows et al. 1997).
L dwarfs should be well represented by the Dusty model, while T dwarfs should be
closer to the Settled model. The $K - L^{\prime}$  colors of the L dwarfs agree quite well
with the Dusty model calculations, and those of the T dwarfs with the Settled model.
However neither model agrees with  the observed $K - M^{\prime}$ colors
for the late L dwarf and the two T dwarfs. 
The most likely explanation of the discrepancy would seem to be the
unexpected absorption by CO in the M bandpass; more 5~$\mu$m spectra are required to 
confirm this.  Such measurements will be more difficult than earlier
anticipated as the 5~$\mu$m flux is around a factor of three smaller than predicted.

\acknowledgments
We are very grateful to the staff at UKIRT, operated by
the Joint Astronomy Centre on behalf of the U.~K.\ Particle
Physics and Astronomy Research Council.
XF acknowledges support from NSF grant PHY-0070928 and a Frank and Peggy
Taplin Fellowship.  DAG acknowledges support from the Center for
Astrophysical Sciences at John Hopkins  University.
The Sloan Digital Sky Survey (SDSS) is a joint project of The
University of Chicago, Fermilab, the Institute for Advanced Study, the
Japan Participation Group, The Johns Hopkins University, the
Max-Planck-Institute for Astronomy (MPIA), the Max-Planck-Institute
for Astrophysics (MPA), New Mexico State University, Princeton
University, the United States Naval Observatory, and the University of
Washington.
Funding for the project has been provided by the Alfred P. Sloan
Foundation, the SDSS member institutions, the National Aeronautics and
Space Administration, the National Science Foundation, the
U.S. Department of Energy, the Japanese Monbukagakusho, and the Max
Planck Society.

\end{document}